\documentclass[twocolumn,showpacs,preprintnumbers,amsmath,amssymb]{revtex4-1}

\usepackage{graphicx}
\usepackage{dcolumn}
\usepackage{bm}

\begin{document}

\preprint{APS/123-QED}

\title{Conception of main local-strong-nonequilibriun state}

\author{Leonid S. Metlov}

 \email{lsmet@fti.dn.ua}
\affiliation{Donetsk Institute of Physics and Engineering, Ukrainian
Academy of Sciences,
\\83114, R.Luxemburg str. 72, Donetsk, Ukraine
}%

\date{\today}

\begin{abstract}
A new conception of main local-strong-nonequilibrium state is proposed on the basis of development 
of the conception local equilibrium state. At list the conception is intended for description of evolution 
of internal defect structure in solids subjected severe external influences, for example, severe plastic 
deformation, radiation and so on, but can be developed on other closely related phenomena. 
\end{abstract}

\maketitle

\section{Analysis of basic conceptions of non-equilibrium thermodynamics}

Plenty of works on non-equilibrium thermodynamics is performed by specialists in mechanics and mathematicians
\cite{cg67,pb69,d70,k05}. Idea, to enter an internal state variable, was offered by Duhem in 1903 yet, and was 
developed at determination of dispersion and attenuation of sound in polyatomic gases \cite{hr28}. Complemented 
by elements of rational mechanics the idea is taking the completed and deductively closed kind in works of 
Coleman and Gurtin \cite{cg67}. In a monograph \cite{pb69} the idea of the internal state variables got further 
development. Here at formulation of thermodynamics of continuous medium the concept of non-equilibrium entropy 
is entered for the first time.

Great importance for development of non-equilibrium thermodynamics is a formulation of it in a language of the 
field theory \cite{d70,k05,gp73}. In a local or substantial form the evolution of a thermodynamic variable $A$ 
is regulated by a balance equation
  \begin{equation}\label{b1}
\dfrac{\partial(a)}{\partial t}=-\vec{\bigtriangledown} \vec{I}_{a,n}+\sigma_{a},
  \end{equation}
where $a$ is the density of an extensive variable $A$, $\vec{I}_{a,n}=a\vec{v}_{a}+\vec{I}_{a}$ is the density 
of complete flux of the field variable $A$, consisting of convective $a\vec{v}_{a}$ and conductive $\vec{I}_{a}$ 
parts, $\vec{v}_{a}$ is rate of transfer of quantity $A$ (barycentric speed), $\sigma_{a}$ is the density of 
internal source power for the same quantity. For the saved quantities the internal sources are absent $\sigma_{a}=0$, 
for other, for example, for entropy, they are calculated from the complete set of constitutive equations.

The variant of non-equilibrium thermodynamics offered by Landau occupies a particular place in the sequence 
\cite{lk54,l37a,l37b}. All qualitative theory of phase transitions was built on a base of the internal state 
variable entered by him in a form of the order parameter. In modern science this direction is continued in phase 
field theories \cite{akv00,esrc02,lpl02,rbkd04,gptwd05,akegny06,rjm09,cn12}.

In obedience to kinetic conception, the behavior of real solids depends not only on average values of 
thermal-motion parameters, but also on them departure from average values, that is, on fluctuations, 
especially, on high-energy fluctuations \cite{rst74}. In many cases, an account of fluctuations is principal 
important, and it determines the level of researches in the area of non-equilibrium processes.

At the same time, besides fluctuation aspect, the problem of non-equilibrium rebuilding of structure of a solid 
has also a force aspect. Fluctuations are important for slow processes, when there is enough time for appearance 
of a large fluctuation. In the case of rapid (intensive) processes, which are mainly examined here, large scale 
thermal fluctuations do not have time to rise up, and it is possible to carry out consideration of such processes 
in the first approaching within the framework of the main field theory.

Internal processes in a solid present a complex picture of mutual transitions and transformations of energy, 
although their nature not always is specified \cite{b64}. Dissipation of energy with participation of defects has 
dynamic nature and is accompanied by transition phenomena of acoustic-like emission. Kinetic energy, arising 
up during defect formation, at first radiates in a form of low-frequency waves, which can be examined, as part 
of a general relaxation process \cite{prtz07}. Scattering of these waves by nonlinear equilibrium thermal 
vibrations completes the relaxation process with participation of defects. Thus, into a chaotic thermal form 
the energy transforms not immediately, but passing the consequence of prior stages.

In the theory of the non-equilibrium phenomena there are two alternative approaches, it is Gibbs statistical 
mechanics, and Einstein statistical thermodynamics \cite{rs00}. In the Gibbs approach the macroscopic laws of 
thermodynamics are deduced by means of averaging of micro-parameters of statistical ensembles, taken in the same 
moment of time \cite{g1902}. In the Einstein approach with the same purpose averaging of fluctuating extensive and 
intensive macro-parameters is carried out in time for one and the same system \cite{rs00}. The identity of these 
results must be guarantee by ergodic hypothesis which is not proven in a general case for present.

In the same time, there is another possibility for ground of thermodynamic laws, it is performed by averaging 
over time in framework of one and the same system, but not on fluctuating macro-parameters (as for Einstein), 
as on the micro-parameters of the system, but not on a statistical ensemble (as for Gibbs). Such possibility 
was applied to the ground for introduction of non-equilibrium temperature and entropy in Ref. \cite{m10} by means of 
averaging over time of motion of separate atoms. Indeed, the macro-parameters of the system in fact already 
in itself are a result of averaging over the whole system, and many subtle features of the system microstates 
are already lost in their remaining fluctuations.

\section{Conception of a local-equilibrium state}

A special place among theories is occupied the Prigogine's self-organization one \cite{gp73}. The theory of 
self-organization is based on principles of local equilibrium. In its foundation a thesis is underlined that 
a global thermodynamic equilibrium over the whole system comes considerably later, than local thermodynamic 
one in different its parts. It is known that in general case the combined first and second law of thermodynamics 
takes a form of an inequality (page 77 \cite{b64})
  \begin{equation}\label{b2}
dU \leq TdS-\sum_{i}F_{i}dA_{i},
  \end{equation}
where $A_{i}$ is an external parameter, $F_{i}$ is the generalized force conjugated to the external parameter. 
In the case of simple bodies $A = V$, $F = p$. Equal sign relates to equilibrium elementary processes, and 
inequality sign relates to the non-equilibrium ones. Only in the case of equilibrium processes a temperature 
and generalized forces can be calculated as partial derivative of internal energy, for non-equilibrium processes 
connection between these quantities and state variables is more complex, depending on the nature of internal processes.

Internal processes can be subdivided into two large classes. It is processes, current with the change of the 
structural state and dissipative processes of type of viscous friction, motions of structural defects et cetera. 
Structural transformations related to the change of component composition as a result of running of chemical 
reactions, formation of structural defects et cetera are strongly non-equilibrium on nature. For their 
description it is impossible to be limited to approaching of local equilibrium, and it is necessary to 
modernize base relation (\ref{b2}).

The most known classic example of such modernization is generalization of relation (\ref{b2}) in case of
multi-component systems with changing number of particles of every type (page 114 \cite{b64})
  \begin{equation}\label{b3}
dU \leq TdS-\sum_{i}F_{i}dA_{i}+\sum_{k}\mu_{k}dN_{k},
  \end{equation}
where $N_{k}$ is number of particles of this sort $k$, $\mu_{k}$ is chemical potential of the component. 
The number of particles of this sort can change both as a result of external influence due to addition or 
withdrawal of components from the system and as result of internal transformations of one component in another.

The second class of internal processes at relatively low power of external influence or in the certain range 
of such influence can be described within the framework of local equilibrium. For a «simple» body relation 
(\ref{b3}) can be written in as equality  
  \begin{equation}\label{b4}
dU = TdS-pdV+\sum_{k}\mu_{k}dN_{k},
  \end{equation}

This relation can be solved with regard to a differential any of variables  $dS$, $dV$ or $dN_{k}$,  and to 
treat these variables as some new thermodynamic potentials. It would allow to express the changes of these 
variables through the changes of all remaining variables, including the internal energy.

Solving relation (\ref{b4}) with respect to entropy, we get possibility, to express its change through 
the changes of measurands, and to calculate it
  \begin{equation}\label{b5}
dS = \dfrac{1}{T}dU+\dfrac{p}{T}dV-\sum_{k}\dfrac{\mu_{k}}{T}dN_{k},
  \end{equation}
or in terms of densities of extensive variables
  \begin{equation}\label{b6}
ds = \dfrac{1}{T}du+\dfrac{p}{T}dv-\sum_{k}\dfrac{\mu_{k}}{T}dn_{k},
  \end{equation}

Speed of entropy production in an adiabatic system in unit of volume is equal
  \begin{equation}\label{b7}
\sigma = \dfrac{ds}{dt} = \dfrac{\partial s}{\partial u} \dfrac{du}{dt}+\dfrac{\partial s}{\partial v} \dfrac{dv}{dt}-
\sum_{k}\dfrac{\partial s}{\partial n_{k}} \dfrac{dn_{k}}{dt}=\sum_{i}I_{i}X_{i},
  \end{equation}  
where  
  \begin{equation}\label{b8}
X_{1} = \dfrac{1}{T},
\quad X_{2} = \dfrac{p}{T},
\quad X_{2+k} = \dfrac{\mu_{k}}{T}
\quad -
  \end{equation}  
are thermodynamic forces (page 257 \cite{b64}), and  
  \begin{equation}\label{b9}
I_{1} = \dfrac{du}{dt},
\quad I_{2} = \dfrac{dv}{dt},
\quad I_{2+k} = \dfrac{dn_{k}}{dt}
\quad -
  \end{equation} 
are thermodynamic flows. Every density of an extensive quantity a for macroscopic system is subjected 
to equation of balance (\ref{b1}).

For finding of explicit expressions for thermodynamic flows and forces it is necessary to invoke the equation 
of balance (\ref{b1}) and the Gibbs relation (\ref{b6}). As an example it is possible to cite the explicit 
expressions for a solid with the gradient of temperature, if to ignore the change of its volume and to eliminate 
the flow of particles (page 259 \cite{b64})
  \begin{equation}\label{b10}
\vec{I}_{s} = \dfrac{1}{T}\vec{I}_{Q},
\quad \vec{X}_{s} = \vec{\triangledown} (\dfrac{1}{T})=-\dfrac{1}{T^2} \vec{\triangledown} T.
  \end{equation}

Here $\vec{I}_{s}$ is the entropy flow, $\vec{I}_{Q}$ is the heat flow. Really thermodynamics force in a form 
(\ref{b10}) appears equal to the gradient of thermodynamics force in a form (\ref{b8}), which in this context it is 
possible to examine as potential of a sort.  
  
At small deviations from the equilibrium state connection between flows and thermodynamic forces is linear 
(page 263 [\cite{b64}])  
  \begin{equation}\label{b11}
I_{i} = \sum_{k}^nL_{ik}X_{k},
  \end{equation}
where $L_{ik}$ is a matrix of kinetic coefficients, the diagonal elements of which are determined by the conjugate 
phenomena of transfer, and the non-diagonal – cross or coupled phenomena. From the invariance of microscopic equations 
in relation to the change of sign with respect to time Onsager proves, that the matrix of kinetic coefficients is 
symmetric \cite{o31a,o31b}
  \begin{equation}\label{b12}
L_{ik} = L_{ki},
  \end{equation}  
Relation (\ref{b12}) is valid only in the case of processes of identical tensor dimension. Thus production of 
entropy (\ref{b7}) for the isotropic systems it is possible to present as a sum of the independent contributions from 
the processes of different tensor dimension. For example, chemical reactions fall into the group of scalar processes, 
heat conductivity and diffusion fall into the group of vectorial processes, viscous motion and motion of structural 
defects can be related to the group of tensor processes. Thus, cross effects between the flows of different tensor 
dimensions are forbidden by Curie principle \cite{p47}  
  \begin{equation}\label{b13}
L_{ik} = L_{k}\delta_{ik}.
  \end{equation}  

Conception of local equilibrium is quite sufficient to solve a wide circle of problems of self-organization of 
complex systems \cite{gp73}. At the same time, the conception is inapplicable for the analysis of processes, for 
example, of treatment of metals by pressure. In this case the density of structural defects during treatment can 
grow in orders of magnitude, and a process is strong-nonequilibrium in nature. For solving of self-organization 
problem in case of structural defects it is necessary to modernize initially basic thermodynamics inequality, maximally 
to take into account the main channels of dissipation, approaching it to equality, and only then to apply presentations 
of \textquotedblleft weakly\textquotedblright nonequilibrium thermodynamics, but in relation to already modernized, 
actually, to the main strong-nonequilibrium state.  
  
Any physical system with large number of particles, as they interact may be complexly between itself, is not able 
to break the law of conservation of energy. Its total internal energy can change only due to an exchange by energy 
with external bodies, for example, in form of work $\delta A$ and/or heat flow $\delta Q$. Unshakeness of this principle 
is expressed by the first law of thermodynamics  
  \begin{equation}\label{b14}
dU = \delta A+\delta Q.
  \end{equation}

Only external influences in one form or other can change internal energy of the system. Internal processes passing 
in the system can only transform one type of energy in another, but does not change its total quantity. 

Those systems, for which two types of exchange by energy with the external systems, thermal and mechanical ones are only, 
fall into the simple systems. Those systems, which besides indicated above channels of exchange by energy with the outer 
world, have also other channels, for example, radiation-damages, electromagnetic or gravity fields et cetera, fall into 
complex systems. In addition, for them it is possible an exchange with the outer world by mass, components and information. 

The simple systems play an important role in development of general principles of classical thermodynamics, and are a 
good starting ground for study of the problem as a whole.

Note, the increment of internal energy in expression (\ref{b14}) has only a form of perfect differential, and for 
equilibrium processes it can be expressed as linear combination of differentials of independent variables of state 
(for the simple systems it is the volume $V$ and entropy $S$)
  \begin{equation}\label{b15}
dU = -pdV+TdS.
  \end{equation}.
  
From where follows $\delta A = – pdV$ and $\delta Q = TdS$, and with definition of perfect differential  
  \begin{equation}\label{b16}
p = -\dfrac{\partial U}{\partial V},
\quad T = \dfrac{\partial U}{\partial S},
  \end{equation}  
where integrating multipliers $p$ (pressure) and $T$ (temperature). If analytical expression for the internal
energy is known, they are determined by simple differentiation.

Relation (\ref{b14}) is just for infinitely slow external influences, when fast-passing internal processes at 
every moment time have had an opportunity to bring a system to the equilibrium state. If external influences 
act with noticeable speed, internal processes not have time to die out, and they must be taken into account 
in total balance of energy transformation. 

The response of solids of the most different nature on strong external mechanical influences is accompanied 
the processes of destruction and fragmentation. The severe loading on metals is accompanied sharp growth of 
number of structural defects – dislocations and grain boundaries, those results in forming of fine-grain structure. 
Generation of structural defects is one of types of internal processes, thus in the process of deformation the number 
of defects is increased on one are two orders. Therefore processes of fragmentation of metals from point of 
thermodynamics is strongly non-equilibrium, and for their description it is impossible to be limited to approaching 
of the local-equilibrium states, and it is necessary to develop the special methods.

\section{Combined first and second law of thermodynamics}

Basic law of thermodynamics for simple solids it is possible to write down in a form
  \begin{equation}\label{b17}
du = \delta a+\delta q=\sigma_{ij}\delta \varepsilon_{ij}+T\delta s',
  \end{equation}
where $u$ is the density of internal energy; $\delta a$, $\delta q$ are work of external forces and increment of heat, 
being on unit of volume; $\sigma_{ij}$, $\varepsilon_{ij}$ are tensors of stresses and deformations; $T$, $s'$ are the 
temperature and entropy density. Here increment of entropy due only to external sources.

A basic law in a form (\ref{b17}) is just both for equilibrium (reversible) and for non-equilibrium (irreversible) 
processes. In the case of equilibrium processes the tensor of deformation have elastic nature 
$\varepsilon_{ij}=\varepsilon_{ij}^e$, and the increment of entropy from outsourcings fully coincides with total 
increment of entropy $\delta s=\delta s'$ (the own sources of entropy are absent). In this case the increment of 
these variables can be equated with perfect differentials
  \begin{equation}\label{b18}
du = \sigma_{ij}d\varepsilon_{ij}+Tds,
  \end{equation}
and the internal energy is the one-to-one function of its arguments $u=u(\varepsilon_{ij},s)$. Variables 
$\varepsilon_{ij}$, $s$ fully and uniquely determine the thermodynamic state of a system in reversible processes.

In the case of irreversible processes part of work goes to the internal heating-up, and work of external forces 
can be divided on reversible and irreversible parts
  \begin{equation}\label{b19}
du = \sigma_{ij}d\varepsilon_{ij}^e+\sigma_{ij} \delta \varepsilon_{ij}^n+T \delta s',
  \end{equation}

We take into account here that elastic deformations are reversible in nature, therefore their increment is 
written in as perfect differential. The stress on the elastic and inelastic element of medium here considered 
identical (body of Maxwell). For dislocations this condition is executed \cite{s79}, and an issue, whether there 
is it executed for other types of defects, remains while opened. 

Further, irreversible part of work of external stresses in Ref. \cite{b64} (second term) goes to formation of 
objects with an excess of potential energy – phonons, grain boundaries, microcracks et cetera, which determine 
the channels of energy dissipation.  If to consider that the thermal channel of dissipation is only present, all 
of irreversible part of work transforms into a heat, and it is necessary to consider
  \begin{equation}\label{b20}
\sigma_{ij} \delta \varepsilon_{ij}^n\equiv T\delta s'',
  \end{equation}
where $\delta s''$ is an additional production of entropy due to internal processes. Combining the production of 
entropy from external and internal sources, we get a thermodynamic identity, which unites the first and second laws 
of thermodynamics for simple bodies
  \begin{equation}\label{b21}
du = \sigma_{ij}d\varepsilon_{ij}^e+T \delta s'+T \delta s''=\sigma_{ij}d\varepsilon_{ij}^e+T ds.
  \end{equation}

Here consider that an additional heat, arising up due to internal sources, had time to come in the equilibrium state, 
therefore for both thermal contributions the same equilibrium temperature is used. In addition, each the type of the 
entropy from external or internal sources does not determine the state of a solid on a separateness; therefore their 
increments are written in through δ-symbol. As memory about the origin of the entropy in a thermal equilibrium state 
is lost, the total entropy in the equilibrium state uniquely determines the state of the system in essence, therefore 
for the record of its increment the sign of perfect differential is used too.

Ordinary thermodynamic inequality, combining the first and second law of thermodynamics, turns out, if irreversible 
part of work, which, actually, already taken into account in a thermal term, once again to add to the identity (\ref{b21})
  \begin{equation}\label{b22}
du \leq \sigma_{ij}d\varepsilon_{ij}+T ds.
  \end{equation}

If not all of heat, arising up due to internal sources, had time to come in the equilibrium state, part of it, is 
in the system in a non-equilibrium form, for example, in a form of non-equilibrium phonons. For this part we have 
not the right to use the equilibrium temperature as a multiplier, guaranteeing implementation of the law of 
conservation of energy over the total system. Therefore for non-equilibrium part of thermo-motion we enter an 
additional term, so that relation (\ref{b21}) is accepted a form
  \begin{equation}\label{b23}
du = \sigma_{ij}d\varepsilon_{ij}^e+T ds+\tilde{T} \delta \tilde{s}.
  \end{equation}

Non-equilibrium temperature $\tilde{T}$ determined from those considering that total energy balance is obeyed 
for the system. Non-equilibrium entropy $\tilde{s}$ is entered phenomenologically. 

A non-equilibrium heat can be in the system either in form of thermal fluctuations or due to the generation of 
non-equilibrium phonons as a result of formation and motion of defects. In first case this quantity is small, 
and for the large systems it is possible to ignore it. In second case it is present as an intermediate product 
of relaxation. As a result of production and motion of defects at severe external influence it is constantly 
generated evenly in all of volume, but in a result of nonlinear dispersion on equilibrium thermal phonons and on 
lattice heterogeneities their energy is constantly pumped over into an equilibrium thermal subsystem. On this 
account a non-equilibrium heat, unlike equilibrium one, cannot accumulate too much in the system, and it always 
considerably less than equilibrium one.

Thus, thermodynamic inequality in a form (\ref{b22}), combining the first and second laws of thermodynamics, 
is just waning expression of general thermodynamic identity (\ref{b21}). Inequality turns out when in Eq. \ref{b22} 
total work made over the system and total heat got from external bodies and made into the system is taken into 
account simultaneously, and transformation of energy on the internal degrees of freedom (\ref{b20}) is not 
taken into account.

If irreversible work goes not only to the heat-up but also on formation of other high-energy objects of type 
of different sort of structural defects, for example, of additional grain boundaries at severe plastic deformation, 
that identity (\ref{b20}) for the special case of account of only one type of defect it is necessary to rewrite in 
a form \cite{m07}
  \begin{equation}\label{b24}
\sigma_{ij} \delta \varepsilon_{ij}^n\equiv T\delta s''+\varphi \delta h,
  \end{equation}
where $\varphi$ and $h$ are the conjugated pair of thermodynamic variables, which describes defectiveness of material, 
that is average energy of a defect and its density. Comparing between itself terms in right side (\ref{b24}), it is 
possible to associate variable $\varphi$ with some static temperature unlike the ordinary thermal temperature $T$, 
having dynamic nature, and variable $h$ – with static entropy. Thus, the thermodynamic identity for a body with one 
type of defects looks like
  \begin{equation}\label{b25}
du = \sigma_{ij}d\varepsilon_{ij}^e+T ds+\tilde{T} \delta \tilde{s}+\varphi \delta h.
  \end{equation}  
  
This record expresses the combined 1st and 2nd law of thermodynamics in form equality (or identity). It is properly 
approximated relationship, as takes into account not all of the possible dissipation channels, but some basic ones only. 
Unlike it the classic expression (\ref{b22}) sets the absolutely exact boundary between reversible and irreversible 
processes, but it talks nothing about their ratio, and in this context on essence of inequality is yet more approximate. 
Vice versa, the relation (\ref{b25}) sets such ratio, and at the correct choice of the basic channels of dissipation it 
can be enough exact.

So, increments of equilibrium thermodynamic variables $\varepsilon_{ij}^e$ and $s$ in (\ref{b25}) expressed through 
perfect differentials, and conjugated them the dual variables $\sigma_{ij}$ and $T$ are calculated by simple 
differentiation of the internal energy
  \begin{equation}\label{b26}
\sigma_{ij} =\dfrac{\partial u}{\partial \varepsilon_{ij}^e},
\quad T = \dfrac{\partial u}{\partial s},
  \end{equation}
that, actually, with the formulas of equilibrium thermodynamics, and, consequently, in accordance with principle of 
local equilibrium.

Increments of non-equilibrium variables $\tilde{s}$ and $h$ is not expressed through perfect differentials, and 
conjugated them dual variables $\tilde{T}$ and $\varphi$ can not be directly calculated by differentiation of the 
internal energy, and for them it is necessary to search the own relations and equations.

As such equations we can use Landau-Khalatnikov-like evolution equations, but written in terms of the internal energy
\begin{eqnarray}\label{b27}
\dfrac{\partial \tilde{s}}{\partial t}=\gamma_{\tilde{s}}(\dfrac{\partial u}{\partial \tilde{s}}-\tilde{T}_{s})
=\gamma_{\tilde{s}}(\tilde{T}-\tilde{T}_{s}),  \\
\dfrac{\partial h}{\partial t}=\gamma_{h}(\dfrac{\partial u}{\partial h}-\varphi_{s})=\gamma_{h}(\varphi-\varphi_{s}).
\end{eqnarray}

Here derivates from internal energy on non-equilibrium variables are essentially the generalized thermodynamic 
forces, and $\tilde{T}_{s}$ and $\varphi_{s}$ are their values in a steady-state. Thermodynamics forces coincide 
with the current values of dual thermodynamic variables, which are included in a thermodynamics «identity» (\ref{b25})
  \begin{equation}\label{b29}
\tilde{T} =\dfrac{\partial u}{\partial \tilde{s}},
\quad \varphi = \dfrac{\partial u}{\partial h}.
  \end{equation}

It is possible to treat variable $\tilde{T}$ as a non-equilibrium temperature, which unlike equilibrium one can take 
on both positive and negative values. Variable $\varphi$ can be understudied as non-equilibrium energy of defect, 
which also can take on both positive and negative values, but in steady-states must be only positive (observable) 
quantity.

\section{Conception of a main local-strong-nonequilibrium state}

Conception of the local-equilibrium state is applicable for global weak-nonequilibrium processes. In this case 
basic thermodynamic relationship takes the form of equality (\ref{b4}). In the case of strong-nonequilibrium 
processes it takes a shape inequality of type (\ref{b3}). There are plenty of attempts to give to it the shape 
of equality by means of formal addition to him of some additional term \cite{e99,etkin1,v01,v02}. Above this 
procedure was realized by means of obvious account of dissipation channels related to the generation of structural 
defects.

On the example of solids with vacancies it is possible to show that relations of type (\ref{b29}) get out not 
arbitrarily, as here, but naturally ensue from procedure of search of the most probable state \cite{m11}[33]. 
It gives a right to present the increments in relation (\ref{b25}) in the form of perfect differentials
  \begin{equation}\label{b30}
du = \sigma_{ij}d\varepsilon_{ij}^e+T ds+\tilde{T} d \tilde{s}+\varphi d h.
  \end{equation}

But then the internal energy is one-to-one function of all its arguments
  \begin{equation}\label{b31}
u = u(\varepsilon_{ij}^e,s, \tilde{s}, h),
  \end{equation}
that, in formal sense all of variables determining the thermodynamics state are equivalent in rights.

Thus, in the case of irreversible processes the state of a system expressed only in equilibrium variables 
$\varepsilon_{ij}^e$ and $s$ is an ambiguous function of these variables. However, extended due to introduction 
of non-equilibrium state variables $\tilde{s}$ and $h$, it becomes the one-to-one function of all arguments.

Relation (\ref{b30}) is generalization of Gibbs equation in case of strong-nonequilibrium processes with 
defect generation in solids. Thus in the classics such relation for strong-nonequilibrium systems carries the 
form of inequality (see of page 77 in \cite{b64}), but the obvious account of the basic channels of energy 
dissipation allowed to reduce the degree of uncertainty and write down this relation in a form of equality or 
identity. Thus, parallel with strong-nonequilibrium processes in a solid, for example, with the generation of 
structural defects, can pass weak-nonequilibrium processes too, for example, viscous motion of structural defects. 
Effect of growth of entropy due to these weak-nonequilibrium processes it is possible to take additionally account 
already from principles of the main local strong-nonequilibrium state, described by equation (\ref{b30}). In this 
case like that, how it is done in the case of the local-equilibrium states, equation must be settled in relation 
to entropy
  \begin{equation}\label{b32}
ds = \dfrac{1}{T}du-\dfrac{\sigma_{ij}}{T}d \varepsilon_{ij}^e-\dfrac{\tilde{T}}{T}d \tilde{s}-\dfrac{\varphi}{T}d h,
  \end{equation}
and expression for the production of entropy can be generalized in a form
  \begin{equation}\label{b33}
\sigma = \dfrac{ds}{dt} = \dfrac{1}{T} \dfrac{du}{dt}-\dfrac{\sigma_{ij}}{T} \dfrac{d \varepsilon_{ij}^e}{dt}-
\dfrac{\tilde{T}}{T} \dfrac{d \tilde{s}}{dt}-\dfrac{\varphi}{T} \dfrac{dh}{dt},
  \end{equation}
from which, knowing an evolution each of parameters $u$, $\varepsilon_{ij}^e$, $\tilde{s}$ and $h$, it is possible 
to find the additional production of entropy.

\begin{acknowledgments}
This work is supported by the budget topic 0109U006004 of the NAS of Ukraine.
\end{acknowledgments}

\end{document}